# Plasmonic abilities of gold and silver spherical nanoantennas in terms of size dependent multipolar resonance frequencies and plasmon damping rates.

K. Kolwas and A. Derkachova
*Institute of Physics, Polish Academy of Sciences,*
*Al. Lotników 32/46, 02-668 Warsaw, Poland*
E-mail: Krystyna.Kolwas@ifpan.edu.pl

## Abstract

Absorbing and emitting optical properties of a spherical plasmonic nanoantenna are described in terms of the size dependent resonance frequencies and damping rates of the multipolar surface plasmons (SP). We provide the plasmon size characteristics for gold and silver spherical particles up to the large size retardation regime where the plasmon radiative damping is significant. We underline the role of the radiation damping in comparison with the energy dissipation damping in formation of receiving and transmitting properties of a plasmonic particle. The size dependence of both: the multipolar SP resonance frequencies and corresponding damping rates can be a convenient tool in tailoring the characteristics of plasmonic nanoantennas for given application. Such characteristics enable to control an operation frequency of a plasmonic nanoantenna and to change the operation range from the spectrally broad to spectrally narrow and vice versa. It is also possible to switch between particle receiving (enhanced absorption) and emitting (enhanced scattering) abilities. Changing the polarization geometry of observation it is possible to effectively separate the dipole and the quadrupole plasmon radiation from all the non-plasmonic contributions to the scattered light.

## 1. Introduction

The unique properties of metal nanostructures [1], [2], [3], [4], [5], [6] are known to be due to excitations of collective oscillations of electron surface densities: surface plasmons (SPs). SPs properties form an attractive starting point for emerging research fields and practical applications (Ref. [7], [8] and references therein). Excitation of SPs at optical frequencies, guiding them along the metal-dielectric interfaces, and transferring them back into freely propagating light, are processes of great importance for manipulation and transmission of light at the nanoscale. A non-diffraction limited transfer of light via a linear chain of gold nanospheres [7], [9], [10], [11], [12], [13] and nanowires [14] can be an example for such control. The geometry- and size-dependent properties of nanoparticles [1], [2], [3], [4], [5], [6] have potential applications in nanophotonics, biophotonics [15], microscopy, data storage, sensing [16], [17] biochemistry [18], [19], medicine [20] and spectroscopic measurements, e.g. in the Surface Enhanced Raman Scattering (SERS) technique [21], [22], [23], [24].

Size dependence of the SP resonance frequency of metallic nanoparticles is essential in applications. The spectral response of metallic nanoparticles can be controlled by changing their size and environment. However, for some applications not only the value of the plasmon resonance frequency is important, but also differences in receiving and transmitting abilities of a plasmonic particle of a given size are crucial. Plasmonic particle of proper size can be used as an effective transmitting nanoantenna and can serve as a transitional structure able to receive or/and transmit electromagnetic radiation at optical frequencies.

The experimental data concerning resonance frequencies of the dipole SP for spherical particles of some chosen sizes [1], [25], [26], [27], [28], [29], comes from measuring positions of maxima in spectra of light scattered or/and absorbed by these particles. Usually, experimental data are in quite good agreement with the predictions of Mie theory [30], [31], [32], [33]. However, Mie theory does not deal with the notion of SPs [29]. As far as we know, there are only few studies that treat SP as an intrinsic property of a metallic sphere and that provide rigorous description of the SP parameters as a direct, continuous function of particle size [28], [29], [34]. It is sometimes supposed [35], that the existing theories rather do not allow direct and accurate calculations of the frequencies of the SP resonant modes as a function of particle size.

In this paper, we describe multipolar resonance frequencies and damping rates of the multipolar SPs for gold and silver spherical particles as a function of their radius up to the large size retardation regime where the plasmon radiative damping is significant. We use the self-consistent electromagnetic approach based on the analysis of the SP dispersion relations [28], [34], [36], [37]. We treat a possibility of excitation of SP oscillations and their damping as a basic, intrinsic property of a conducting sphere. We provide some ready-to-use surface multipolar plasmon size characteristics of the dipole and higher polarity plasmon resonances in gold and silver nanospheres up to the radius of $150 nm$ covering the multipolar plasmon resonance frequencies in the range of about $1 \div 4 eV$ for silver and $1.2 \div 2.7 eV$ for gold nanoparticles. Optical properties of these metals are described by realistic, frequency dependent refractive indices of corresponding pure metals [38]. The multipolar SPs resonance frequencies and the SPs damping rates are treated consistently.

Implications of damping processes in SP applications are extremely important ( [4], [39], [40] and references therein). In this paper, we underline the role of the plasmon radiation damping in comparison with the internal energy dissipation damping in formation of receiving and transmitting properties of a plasmonic particle. In particular we demonstrate that the enhancement factor of absorbing and scattering particle abilities depends on the relative contribution of the energy dissipative and radiative processes to the total damping of SP oscillations. These processes are size dependent.

We show that only knowing the size dependence of both: the multipolar SP resonance frequency and corresponding damping rate, an effective description of plasmonic properties is possible. SP size characteristics we study here allow not only to predict a frequency of SP resonance in given polarity order, but also to exploit a plasmonic particle as an effective receiving (enhanced absorption) or emitting (enhanced scattering) nanoantenna. These features are crucial for SPs applications.

## 2. Modelling of SPs inherent size characteristics

Collective motion of surface free-electrons in a metallic particle can be excited by the EM field under the resonance conditions which is defined by intrinsic properties of a plasmonic sphere due to its size, optical (conductive) properties of a metal and dielectric properties of the sphere's environment. The SP resonance takes place when the frequency of the incoming light $\omega$ approaches at least one of the characteristic eigenfrequencies $\omega_l(R)$, $l = 1,2,3,...$ of a sphere of radius $R$. The dynamic plasmon charge density distribution, induced by EM field, can be quite complicated. With the increasing of the sphere surface, the surface charge density distributions of polarity higher than the dipole one ($l = 1$) come into play [34].

The collectively oscillating electrons of SPs at curved surface must emit EM energy through radiation. Radiative damping is, then, the inherent property of SPs. It enables the enhancement of the EM field scattered by a sphere and is inseparably associated with SP oscillations at characteristic frequencies of multipolar plasmon modes.

To find the SP intrinsic size characteristics we use a self-consistent rigorous EM approach based on [36], and described in more details e.g. in [28], [29], [34]. We consider continuity relations at the spherical boundary for the tangent component of the transverse magnetic (TM) solution of the Helmholtz equation in spherical coordinates, while TM solutions only possess nonzero normal to the surface component of the electric field $E_r$. This component is able to couple with the charge densities at the boundary. Resulting conditions define the dispersion relations in spherical coordinates [28], [34]:

$$\sqrt{\varepsilon_{in}(\omega)}\xi_l^{'}(k_{out}(\omega)R)\psi_l(k_{in}(\omega)R) + \sqrt{\varepsilon_{out}(\omega)}\xi_l(k_{out}(\omega)R)\psi_l^{'}(k_{in}(\omega)R) = 0, \quad (1)$$

which are fulfilled for the complex eigenfrequencies of the fields $\Omega_l$, $l = 1,2,3,...$ at the $r = R$ distance from a sphere centre. $\psi_l(z)$ and $\xi_l(z)$ are Riccati-Bessel spherical functions, the prime marker (') indicates differentiation with respect to the function's argument and $k_{in} = \frac{\omega}{c}\sqrt{\varepsilon_{in}(\omega)}$ and $k_{out} = \frac{\omega}{c}\sqrt{\varepsilon_{out}}$.

The realistic modelling of the SP characteristics is possible if the functional dependences of the dielectric function on frequency in analytical form for both the sphere $\varepsilon_{in}(\omega)$ and surrounding medium $\varepsilon_{out}(\omega)$ are known. It assures the correct coupling of the metal dispersion to the overall frequency dependence of the plasmon dispersion relation Eq.(1). We used the modified dielectric function of the Drude electron gas model based on the kinetic gas theory: $\varepsilon_{in}(\omega) = \varepsilon_D(\omega) = \varepsilon_\infty - \omega_p^2/(\omega^2 + i\gamma\omega)$ which quite well describes the optical properties of many metals within relatively wide frequency range [38]. $\varepsilon_\infty$ is the phenomenological parameter describing the contribution of the bound electrons to the polarizability [41] which equals to $1$ only if the conduction band electrons are responsible for the optical properties of a metal (e.g. sodium) [28]. For gold and silver, the interband transitions are important for defining their optical properties. $\omega_p$ is the bulk plasmon frequency, $\gamma$ is the phenomenological relaxation constant of the bulk material. For a perfect free-electron bulk metal with infinite boundaries, electron relaxation is due to electron-electron, electron-phonon, and electron-defect (grain boundaries, impurities, and dislocations) scattering processes. $\gamma$ results from the average of the respective collision frequencies of electrons and thus is closely related to the electrical resistivity of the metal [24]. The functions $\varepsilon_{in}(\omega)$ with effective parameters: $\varepsilon_\infty = 9.84$, $\omega_p = 9.096\,eV$, $\gamma = 0.072 eV$ for gold, and $\varepsilon_\infty = 3.7 eV$, $\omega_p = 8.9 eV$, $\gamma = 0.021 eV$ for silver satisfactorily reproduce the experimental values of $Re[n^2(\omega)](Im[n^2(\omega)])$ in the frequency ranges: $0.8 eV - 5.0 eV$ for gold and $0.8 eV - 4.2(4.0) eV$ for silver (Figure 1). As illustrated in Figure 1, the agreement of the model and measured frequency dependence of the dielectric function is very good, with exception of $Im\varepsilon_{in}(\omega)$ dependence for silver at higher frequencies of the studied frequency range.(above 4eV, it is for some wavelengths outside the visible range($\lambda$<310nm)).

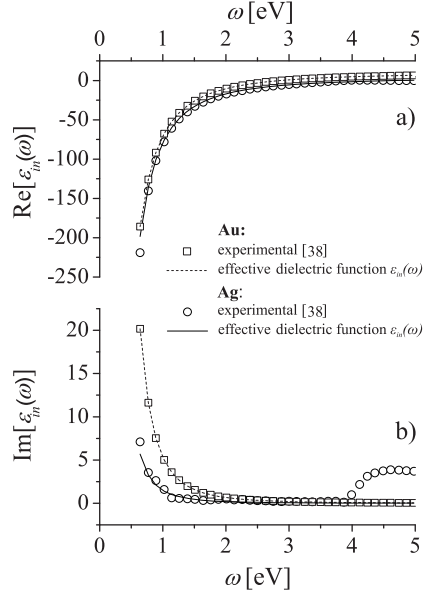

Figure 1: Real and imaginary parts of the dielectric function with effective parameters: $\varepsilon_\infty = 9.84$, $\omega_p = 9.096 eV$, $\gamma = 0.072 eV$ for gold, and $\varepsilon_\infty = 3.7 eV$, $\omega_p = 8.9 eV$, $\gamma = 0.021 eV$ for silver. Squares and circles are the corresponding values of Re($n^2$) and of Im($n^2$) according to [38].

In the calculations the optical properties of the dielectric medium outside the sphere are described by $\varepsilon_{out} = n_{out}^2$, with $n_{out}$ is chosen here to be $1$ or $1.5$. However, the formalism, in general, also includes the frequency dependence of the optical properties of the surroundings. In such a case, the frequency dependence of both the metallic sphere and the surrounding should be convoluted into the size dependence of multipolar plasmon eigenfrequencies and the corresponding plasmon damping rates.

The dispersion relation Eq.(1) is solved numerically with respect to complex values of $\omega = \Omega_l(R)$ for each $l$. They define the size dependent multipolar plasmon oscillation frequencies $\omega'_l(R) = Re\Omega_l(R)$ and the damping rates $\omega''_l(R) = Im\Omega_l(R)$ of plasmonic oscillations. The sphere radius $R$ is treated as an outside, independent parameter. The results are presented in Figure 2 for silver and in Figure 3 for gold spheres. Figures a and c show the plasmon size characteristics for free spheres in vacuum ($n_{out} = 1$), while Figures b and d show the same size dependencies for spheres suspended in a dielectric medium of $n_{out} = 1.5$. The first seven multipolar plasmon characteristics are presented (with $l = 1$ for a dipole plasmon and $l = 2$ for a quadrupole plasmon) up to the radius $150 nm$. The multipolar plasmon resonance frequencies are lower for gold than for silver nanospheres of the same size and with the same surroundings.

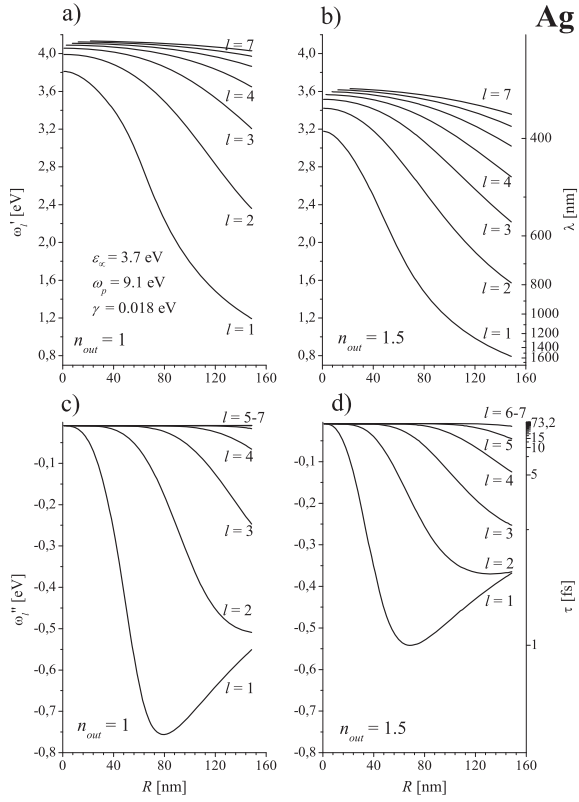
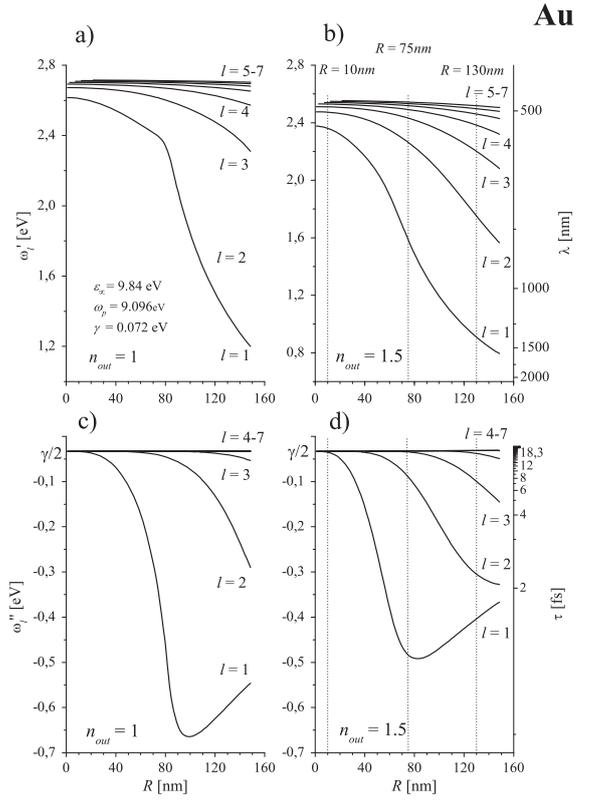

Figure 2: (a), (b) Multipolar plasmon resonance frequencies $\omega'_l(R)$ and (c), (d) plasmon damping rates $\omega''_l(R)$ ($l=1,2,..7$) for silver nanoparticles in free space ($n_{out}=1$) and in a suspension ($n_{out}=1.5$) as a function of sphere radius.

Figure 3: (a),(b) Multipolar plasmon resonance frequencies $\omega'_l(R)$ and (c), (d) plasmon damping rates $\omega''_l(R)$ ($l=1,2,..7$) for gold nanoparticles in free space ($n_{out}=1$) and in a suspension ($n_{out}=1.5$) as a function of sphere radius.

As our numerical results show (Figures 2 and 3 for silver and gold and, [28], [34] for alkalies), the plasmon multipolar resonance frequencies $\omega'_{0,l}$ and the corresponding damping rates $|\omega''_{0,l}|$ (Eq. (6)) can be attributed only to a sphere of a radius larger than minimal radius $R_{\min,l}$ in each plasmon mode. $R_{\min,l}$ is the fast increasing function of $l$.

The extreme values of $\omega'_l(R)$ and $\omega''_l(R)$ can be found from an approximated and very rough consideration using the power series expansion of the spherical Bassel and Hankel functions:

$$j_l(z) = \frac{z^l}{(2l+1)!!}\left[1 - \frac{0.5z^2}{1!(2l+3)} + \frac{(0.5z^2)^2}{2!(2l+3)(2l+5)} - \ldots\right], \qquad (2)$$

$$h_l(z) = -i\frac{(2l-1)!!}{z^{l+1}}\left[1 - \frac{0.5z^2}{1!(1-2l)} + \frac{(0.5z^2)^2}{2!(1-2l)(3-2l)} - \ldots\right], \qquad (3)$$

where $(2l\pm1)!!\equiv 1\times 3\times 5\times\ldots\times(2l\pm1)$. Applying the limit of small arguments z (so called "quasistatic approximation"), and keeping only the first terms of the power series Eqs. (2) and (3) (what is not justified for larger $l$ values), the dispersion relation Eq. (1) is fulfilled under the condition below:

$$-\frac{l}{l+1}\frac{\varepsilon_{in}}{\varepsilon_{out}}=1, \tag{4}$$

For equation $\varepsilon_{in}(\omega)$ in the form given by Eq. (1), one gets:

$$\omega'_{0,l}=\left[\frac{\omega_p^2}{\varepsilon_\infty+\frac{l+1}{l}\varepsilon_{out}}-\left(\frac{\gamma}{2}\right)^2\right]^{1/2}, \tag{5}$$

$$\omega''_{0,l}=-\frac{\gamma}{2}. \tag{6}$$

Neglecting relaxation ($\gamma=0$) for a perfect free-electron metal ($\varepsilon_\infty=1$) and $\varepsilon_{out}=1$ Eq.(5) leads to the well-known multipolar plasmon frequency values of a metal sphere within the "quasistatic approximation" [37], [42], [43]: $\omega'_{0,l}=\omega_p/\sqrt{1+\varepsilon_{out}(l+1)/l}$, and, in particular, to the dipole ($l=1$) plasmon frequency $\omega'_{0,l=1}=\omega_p/\sqrt{3}$, the value known as the giant Mie resonance frequency.

Plasmon oscillations are always damped (see drawings c and d in Figures 2 and 3 due to radiation losses and all the relaxation processes included in the relaxation rate $\gamma$. With increasing size, the damping rates $|\omega''_l(R)|$ initially increase starting from the values $|\omega''_l|\approx|\omega''_{0,l}|=\gamma/2$ in each plasmon mode. If $\gamma$ accounts for electronic relaxation processes leading to dissipation (absorption) of energy in the metal, $|\omega''_{0,l}|$ accounts for the decay of the SP oscillations due to dephasing of the collective electron motion which is often assumed to be statistically "memory destroying". This quantity has been investigated by various experimental methods ( [4], [39], [40] and references therein). For particles of radii larger than $R_{min,l}$, the size dependence of the total damping rates $|\omega''_l(R)|$ is due to the size dependence of the radiative damping rates $\gamma_l^{rad}(R)$. Whereas radiative dumping and energy dissipation are uncorrelated processes, the total damping rate $|\omega''_l(R)|$ can be written as a sum:

$$|\omega''_l(R)|=\gamma_l^{rad}(R)+\gamma^{diss}, \quad l=1,2,3... , \tag{7}$$

where $\gamma^{diss}=\gamma/2$ for any multipolar mode $l$. Size dependence of the rate $|\omega''_l(R)|$ is results from $\gamma_l^{rad}(R)$ size dependence. The initial increase of $|\omega''_l(R)|$ with particle size is followed by the subsequent decrease for sufficiently large spheres as is illustrated in Figures 2c,d and 3c,d for the dipole mode damping $|\omega''_{l=1}(R)|$ for silver and gold nanospheres, respectively.

The optical properties of the surrounding medium can introduce a significant alteration in the multipolar SP resonance frequencies and damping rates for a particle of given radius. In particular, an increase of the optical density of the surroundings introduces an important "red shift" of the plasmon resonance frequencies $\omega'_l(R)$, as is illustrated in Figures 2,3a and b.

## 3. Properties of plasmonic nanoantennas in terms of multipolar SP resonance frequencies and plasmon oscillation damping rates

The derived SP size characteristics can be used to tune the absorbing or emitting properties of the plasmonic spherical nanoantennas. Knowing both the multipolar SP resonance frequencies $\omega'_l(R)$ and corresponding damping rates $|\omega_l^{''}(R)|$, one can predict properties of plasmonic particle in response to the external EM field and find the optimal size of a nanosphere for a chosen application. It can be, for example, an effective coherent coupling with some neighbouring particles in a metal nanoparticle array, an enhancement of the EM field near the sphere surface (used e.g. for SERS spectroscopic technique) or a modification of the far field scattering properties.

Resonant excitation of SP oscillations is possible when the frequency of the incoming light $\omega_{in}$ approaches an eigenfrequency of a plasmonic nanoantenna of given radius $R$: $\omega_{in} = \omega_l^{'}(R)$, $l = 1,2,3,...$. (Figures 2,3a and b). If excited, plasmon oscillations are damped at the corresponding rates $|\omega_l^{''}(R)|$. The sphere acts, then, as a receiving or/and emitting multipolar EM nanoantenna in mode(s) $l$. The nanoantennas performance can be adjusted by changing the sphere size. Not only the resonance frequencies $\omega_l^{'}(R)$ (with $l = 1$ for the dipole antenna mode), but also receiving (absorbing) and scattering (transmitting) abilities of plasmonic nanoantennas are size dependent as follows from the SP damping rates $|\omega_l^{''}(R)|$ size dependence (Figures 2,3 c,d). It is convenient to keep expressions for $\omega_l^{'}$, and $\omega_l^{''}$ in [eV] units in the discussion, but for completeness we give the corresponding magnitudes in [1/s] ones on the left vertical axis of Figures 2,3 c,d.

## 3.1. Receiving and transmitting abilities of plasmonic nanoantenna as an intrinsic property of a plasmonic particle

It is widely known, that optical properties of a perfect-metal spherical particle, which is much smaller than the wavelength of incoming light, are mainly due to the giant absorption at a resonance frequency $\omega_p/\sqrt{3}$. That value corresponds to the SP dipole resonance frequency $\omega_{0,l}^{'}(R)$ given by Eq.(5) for $l = 1$. However, a plasmonic particle can act as an efficient absorbing antenna not only in the dipole, but also in larger polarity modes for particles of larger sizes. Moreover, in some ranges of parameters $R$ and $l$, the receiving and radiative abilities of a plasmonic nanoantenna may be comparable. In such a case the plasmonic fraction in the total extinction spectrum would be manifested by comparable contributions of the absorption and scattering.

Scattering and absorbing abilities of spherical nanoantennas can be conveniently described by the relative contribution of the radiative and nonradiative damping rates to the total damping rate $|\omega_l^{''}(R)|$. (Figures 2,3, a,b). Plasmonic nanoparticles can act as the absorbing antennas when the contribution of the nonradiative damping rate in the total one is large. That is possible for particles of some radii $R$, such that $|\omega_l^{''}(R_{\min,l})| \approx |\omega_{0,l}^{''}| = \gamma^{diss}$. In such case, the SP resonance manifestation in the absorption spectrum is expected to be well pronounced, with maxima at the resonance frequencies $\omega_l^{'}(R)$, $l = 1,2,...$. It also contributes considerably to the extinction spectrum regardless of how large the particle is. On the other hand, if contribution of the radiative damping $\gamma_l^{rad}(R)$ to the total SP damping rate $|\omega_l^{''}(R)|$ is large in comparison with the nonradiative damping $\gamma^{diss}$: $|\omega_l^{''}(R)| \approx \gamma_l^{rad}(R)$, a particle of radius $R$ is able to act as an efficient scattering antenna. The range of particle sizes which fulfil these conditions results from the size characteristics of $|\omega_l^{''}(R)|$ presented in Figure 2 and Figure 3 c,d for silver and gold plasmonic nanospheres, respectively. The basic, intrinsic properties of a particle described by $\omega_l^{'}(R)$ and the size characteristics of $|\omega_l^{''}(R)|$ are reflected in a way the particle responds to the light field.

## 3.2. SP resonance manifestation in optical signals

Let us discuss some examples of light elastically scattered by gold nanoparticles of chosen sizes in environment with $n_{out} = 1,5$. According to Mie theory [30], [31], [32], [33], [44], the optical response to the incoming EM wave can be described in spherical coordinates as square of sum of the partial waves of the TM ("electric") and the TE ("magnetic") EM contributions. Therefore, the fields contributing to the light intensity are inevitably composed of both TM and TE components of different polarities $l$, while the electromagnetic fields of SP dispersion relation (Eq.(1)) are transverse magnetic. Constructive and destructive interference of TM and TE components influences the manner of plasmons manifestation in the scattered light intensity (irradiance). For given particle size, the manifestation of the SP resonance depends on the observable quantity, observation angle and polarization geometry. Figure 4 illustrates the spectra of the total scattering $\sigma_{scat}(\omega)$, extinction $\sigma_{ext}(\omega)$ and absorption $\sigma_{abs}(\omega)$ cross-sections for gold spheres with radius $10nm, 75nm$ and $130nm$ (Figure 4a,b and c, respectively) calculated within Mie theory for a suspension ($n_{out} = 1.5$) of mono-sized gold spheres as an example.

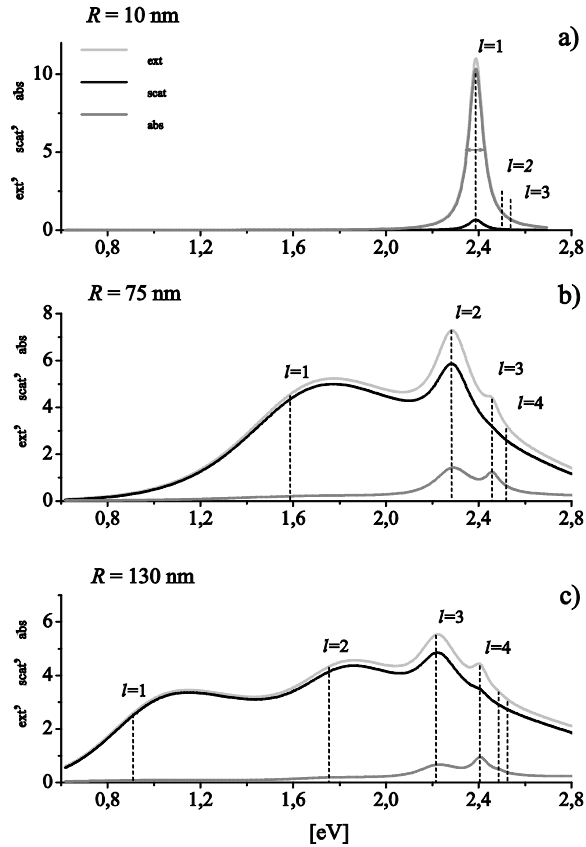

Figure 4: Total extinction, scattering and absorption cross-sections spectra for gold nanospheres of radii: a) $10nm$, b) $75nm$ and c) $130nm$ in suspension ($n_{out} = 1.5$). Dotted vertical lines indicate the SP multipolar resonance frequencies $\omega'_l(R)$, $l = 1,2,3,..$ for these radii, according to the data in Figure 3 b.

For $R = 10 nm$ (Figure 4a), the dipole plasmon resonance at $\omega'_{l=1}(R = 10nm)$ becomes apparent mainly in the enhanced absorptive properties of a nanosphere (the maximum in $\sigma_{abs}(\omega)$ dependence, gray line in Figure 4a), due to the large contribution of the nonradiative damping to the total damping rate: $|\omega''_l(R = 10nm)| \approx \gamma/2 = 0,036 eV$. The on-resonance enhancement factor is of three orders in magnitude. The peak in the total scattering cross-section $\sigma_{scat}(\omega)$ is by a factor of ten smaller due to the poor radiation damping: $|\omega''_{l=1}(R = 10nm)| \geq \gamma^{rad}_l(R)$ (Figure 3d). Therefore, the optical properties (including colours) of such metallic nanoparticles are dominated by the resonant nonradiative plasmon energy dissipation at the dipole SP eigenfrequency $\omega'_{l=1}(R = 10nm)$ (see Figure 3b). The mentioned above also applies to the absorptive properties of atoms or molecules associated with the resonant (dipole) transition to the short-living excited states of lifetimes of the order of picoseconds. The full width at half maximum $\Gamma$ (FWHM) of a narrow, Lorentzian-like peak in the total absorption cross-section $\sigma_{abs}(\omega)$ (Figure 4a) is well described by the plasmon damping rate $|\omega''_{l=1}(R = 10nm)|$ (see Figure 3d):

$$\Gamma(R = 10nm) = 2|\omega''_{l=1}(R = 10nm)| \approx \gamma. \qquad (8)$$

The corresponding plasmon damping time $\tau_\gamma = 18,3 p\sec$.

Scattering effects become important for nanoparticles of larger sizes, as known from Mie work [30]. Figure 4b illustrates this effect for sphere with radius $R = 75 nm$. The magnitude of the total scattering cross-section $\sigma_{scat}(\omega)$ is bigger than that of the total absorption cross-section $\sigma_{abs}(\omega)$ in the full optical range. The scattering spectrum is broadened and composed of some partially overlapping maxima which are blue shifted in respect to the SP multipolar resonance frequencies $\omega'_l(R)$. Using again the SP damping rate characteristics, one can predict the multipolar plasmonic contribution to the scattering.

The collectively oscillating electrons at the resonance frequency $\omega'_l(R = 75nm) = 1,6 eV$ lose their energy mainly due to radiation, the effect accounted in the size augmented radiation rate contribution to the total damping rate: $|\omega''_l(R = 75nm)| = 0,49 eV \approx \gamma^{rad}_l(R = 75nm)$ (Figure 3d). That is why the fraction of the dipole plasmonic absorption is negligibly small. Corresponding dipole damping rate $|\omega''_{l=1}(R = 75nm)| = 0,49 eV$ (Figure 3d) is dominated by the radiation damping: $\gamma^{rad}_l(R = 75nm) \geq \gamma^{diss} = 0,036 eV$. Therefore, if the plasmon oscillations are continuously excited by light at the resonance frequency of a dipole plasmon $\omega = \omega'_{l=1}(R = 75nm)$, a sphere is able to efficiently scatter light through the plasmonic mechanism and to work as an excellent emitting antenna at that frequency. Enhancement of the scattered near and far field in the space around the particle is than possible.

The larger is the rate $\gamma^{rad}_l(R)$ in comparison with the nonradiative damping rate for successive $l = 1,2,3...$, the higher is the peak in the scattering spectrum $\sigma_{scat}(\omega)$ due to the plasmonic contribution, as demonstrated in Figure 4b and c. But also, the higher is the rate $\gamma^{rad}_l(R)$, the larger is the spectral bandwidth of the SP participation in the scattering spectrum around the SP resonance frequency $\omega'_l(R)$.

As illustrated in Figure 4 and Figure 6b, the larger is the spectral bandwidth of the SP contribution of polarity $l$ to the spectrum (defined by the rate $\gamma^{rad}_l(R)$), the stronger is the blue shift of the maximum in respect to the SP resonance frequency $\omega_l^{'}(R)$. As illustrated in Figure 4b,c, all the large-bandwidth maxima in the scattering spectrum suffer from this effect and are blue-shifted in respect to the SP resonance frequencies $\omega_l^{'}(R)$. This effect is due to the important changes with frequency in the imaginary part of the index of refraction $n_{in}(\omega)$ (Figure 6c), that affects the frequency dependence of the nonplasmonic contribution (specular reflection) to the scattering. But also, the larger is the spectral bandwidth of the maximum, the more important is of the interference effects (constructive and destructive) of the partial TM and TE waves.

Even large plasmonic particles (such as those of radius $R = 75nm$ or larger (see Figure 4b,c)) can act as receiving antennas producing narrow, well pronounced peaks in the absorption (and extinction) spectra, if only the participation of the energy dissipation rate $\gamma^{diss}$ in the total plasmon damping rates $\left|\omega_l^{''}(R)\right|$ in mode $l$ is comparable with the radiative rate $\gamma_l^{rad}(R)$: $\gamma_l^{rad}(R) \approx \gamma^{diss}$ (see Eq.(7)). The well pronounced maxima in the total absorption cross-sections $\sigma_{abs}(\omega)$ (black lines in Figure 4) near the plasmon frequencies $\omega_{l=2,3}^{'}(R=75nm)$ (Figure 4b) and $\omega_{l=4}^{'}(R=130nm)$ (Figure 4c) are some examples. The well pronounced absorption peaks due to SPs are not affected by shifting and smearing out effects suffered by maxima in the scattered spectra, while the SP absorption is due to the plasmonic dependent energy dissipation only. Therefore, interference of EM waves does not affect the position and width of maxima in $\sigma_{abs}(\omega)$ (black lines in Figure 4). The contribution of the dipole SP resonance to the maximum in the scattering spectrum of a sphere with $R = 130nm$ is smaller than for a sphere with $R = 75nm$, as one can expect knowing that $\left|\omega_{l=1}^{''}(R=130nm)\right| < \left|\omega_{l=1}^{''}(R=75nm)\right|$ (see Figure 3d). Such relation results from the smaller contribution of the radiative damping in the total SP damping rate: $\gamma_{l=1}^{rad}(R=130nm) < \gamma_{l=1}^{rad}(R=75nm)$.

As was just mentioned, the partial smearing out and blue shift of the maxima in the scattering (extinction) spectra is due to the interference of EM fields scattered by a sphere. Such fields are inevitably composed of both the TM and TE components of different polarity $l$, while the resonant SP's contributions are due to the normal to the sphere surface component of the electric field $E_r$, that is present in the TM polarization mode only. The TE fraction of eddy currents is also size dependent and increases monotonically with $R$, contributing also at some amount to the blue shift of the maxima in the total scattering cross-section [29].

One can conclude, that if the SP damping rates $\left|\omega_{l=1}^{''}(R)\right|$ for $l=1,2,...$ of a plasmon active sphere of radius $R$ are mainly due to the nonradiative damping $\left|\omega_l^{''}(R)\right| \approx \gamma^{diss}$ (Figure 3d), such sphere is able to efficiently absorb light through the plasmonic mechanism near the corresponding SP resonance frequency $\omega_l^{'}(R)$ in spectrally narrow bandwidth $\gamma^{diss}$ defined by the relaxation rate $\gamma$. However, such receiving antennas are rather unable to couple electromagnetically with another particle by the plasmonic mechanism.

If the SP damping rate $\left|\omega_{l=1}^{''}(R)\right|$ are much larger than $\gamma^{diss}$, plasmonic particle of the size $R$ is a good radiating antenna at resonance frequency $\omega_l^{'}(R)$ of given polarity $l$. Radiative damping, which is inherently coupled to plasmon oscillations, affects not only the spectral width of the plasmon related maxima in the scattered light intensity, but also enhances the plasmon related contributions to the scattered spectrum. For large radiation damping rates $\left|\omega_{l=1}^{''}(R)\right|$, the energy of the incident EM wave can be effectively redistributed into the scattered field energy around the sphere due to the plasmonic mechanism (a sphere acts as a radiating antenna).

### 3.3. Plasmonic scattering in orthogonal polarization geometries

A distinct dipole and quadrupole scattering abilities of the plasmonic nanoantenna can be conveniently demonstrated by linearly polarized light illuminating the plasmonic sphere at the right angle to the direction of the incoming light beam (Figure 5).

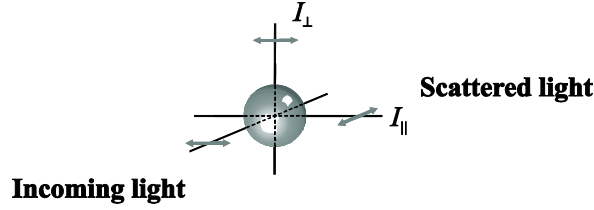

Figure 5: Orthogonal observation geometry

Such "clinical" orthogonal scattering geometry enables to expose a contribution of both: a single dipole ($l=1$) and a single quadrupole ($l=2$) SP to the scattering signals and to separate these contributions spatially by observing $I_\perp(\omega, R)$ and $I_\parallel(\omega, R)$ intensities (Figure 4) [29], [45]. Whereas $I_\perp$ and $I_\parallel$ are the intensities of the purely scattered light while there is no influence of the interference effect with the EM field of the incoming light wave from the perpendicular direction. Therefore, such observation enables to study pure scattering abilities of a plasmonic sphere. Using such an experimental geometry and observing the changes in $I_\perp(R)/2\pi R^2$ and $I_\parallel(R)/2\pi R^2$ with size [46] we have studied the size dependence of the dipole and quadrupole plasmon resonances in our experiment on spontaneously growing sodium droplets (up to the droplet radius $R=150 nm$) induced by laser light [28], [29]. Studying the intensity scattered by the particle unit area allowed us to diminish a contribution of the background due to eddy currents and to emphasize SP manifestation.

The changes in the spectra of $I_\perp(\omega)$ and $I_\parallel(\omega)$ with particle radius, discussed in terms of the intrinsic SP multipolar size characteristics (Section 2), can directly illustrate the role of the SP radiative damping rates $\gamma_l^{''}(R)$ in formation of the radiative properties of the plasmonic nanoantennas.

The FWHM $\Gamma_{l=1}^{scat}$ of the $I_\perp(\omega)$ spectrum (Figure 6 a,b, solid line), and $\Gamma_{l=2}^{scatt}$ of the $I_\parallel(\omega)$ spectrum (Figure 6 a,b, dashed line) can be described by the Lorentzian function of FWHM corresponding values of the plasmon damping rates $\omega_{l=1}^{''}(R)$ and $\omega_{l=2}^{''}(R)$:

$$\Gamma_l^{scatt}(R) \approx 2\left|\omega_l^{''}(R)\right|, \quad l=1,2. \qquad (9)$$

The peaks in the spectra for spheres of larger sizes become blue shifted in respect to the intrinsic value $\omega_{l=1,2}^{'}(R)$ of the SP dipole and quadrupole SP resonances, as illustrated in Figure 6 a and c. The larger the spectral width $\Gamma_l^{scatt}$, the stronger the blue shift of the maximum position $\omega_l^{max}$ in respect to the SP plasmon resonance position $\omega^{'}(R_l)$, as illustrated in Figure 6 b. The maxima due to SP resonances in the scattering spectra are governed by the radiative damping rates $\gamma_l^{rad} = \left|\omega_l^{''}(R_l)\right| - \gamma^{diss}$, for dipole ($l=1$) and quadrupole ($l=2$) plasmon mode, correspondingly.

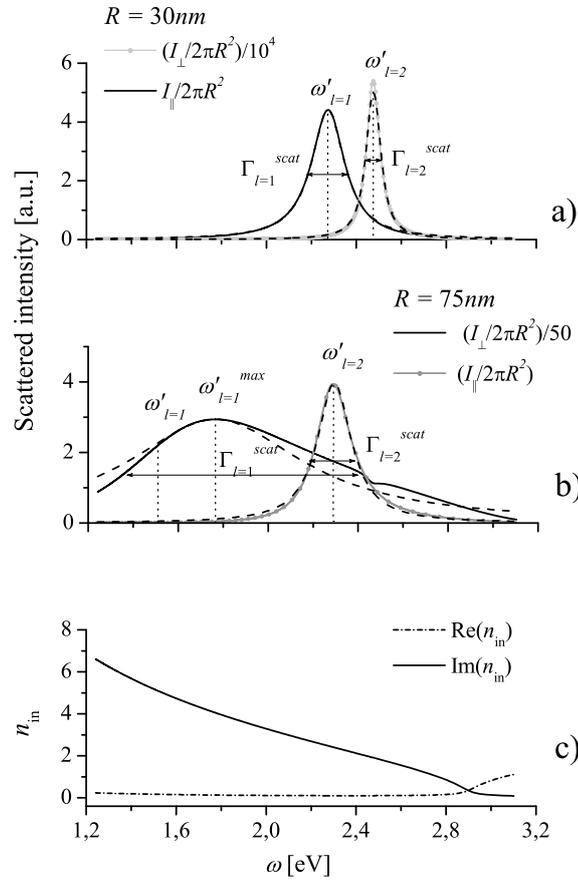

Figure 6: a) and b) $I_\perp(\omega)$ and $I_\parallel(\omega)$ spectra for $R = 30nm$ and $R = 75nm$, respectively, with the SP dipole resonance at $\omega = \omega'_{l=1}(R)$ and the quadrupole resonance at $\omega = \omega'_{l=2}(R)$, correspondingly. Dashed lines represent Lorentz profiles of FWHM $\Gamma^{scat}_{l=1}$ Eq. (9). c) The dependence of the refraction coefficient ($Re$ and $Im$ part) of the bulk gold on $\omega$ in the studied frequency range. $n_{out} = 1.5$.

One can conclude that changing the polarization geometry it is possible to separate the quadrupole plasmon radiation from the dipole and from the larger multipolarity plasmon contribution to the scattered light. Knowing the manner of dipole and quadrupole plasmon manifestation in $I_\perp(\omega)$ and $I_\parallel(\omega)$ spectra correspondingly, and $\omega'_{l=1,2}(R)$ and $\omega''_{l=1,2}(R)$ size dependencies, it is possible to switch between the receiving (enhanced absorption) and emitting (enhanced scattering) abilities of a plasmonic nanoantenna by changing the direction of observation (Figure 5) or direction of polarization in respect to the observation plane. It is also possible to control a bandwidth of particle extinction or scattering spectra changing it from spectrally broad to spectrally narrow and vice versa, by changing the wavelength (see Figure 4b or Figure 6b as an example) of the of illuminating light, if a particle is sufficiently large.

## 4. Conclusions

Receiving or/and emitting properties of the spherical plasmonic nanoantenna are governed by the inherent parameters: the size dependent plasmon resonance frequencies $\omega'_l(R)$ and damping rates $\omega''_l(R) = \gamma_l^{rad}(R) + \gamma^{diss}$, which provide a complete description of particle plasmonic properties. Knowledge of the size dependence of both: the multipolar SP resonance frequencies and corresponding damping rates makes possible to effectively describe particle plasmonic properties in response to the EM field. Taking advantage of the particle intrinsic size characteristics it is possible to predict not only the SP resonance frequencies $\omega'_l(R)$ but also the strength of SP resonances in given polarity order $l = 1,2,3,...$ after proper adjustment of the relative magnitude of the radiative $\gamma_l^{rad}(R)$ and energy dissipative $\gamma^{diss}$ damping rates.

If the SP multipolar damping rate $|\omega''_{l=1}(R)|$ of a plasmon active sphere of radius $R$ is mainly due to the nonradiative rate: $|\omega''_l(R)| \approx \gamma^{diss} = \gamma/2$ (Figure 3d), such sphere is able to efficiently absorb light through the plasmonic mechanism near the corresponding SP resonance frequency $\omega'_l(R)$, in spectrally narrow bandwidth $\gamma^{diss}$, even for larger sizes (and larger $l$). However, such receiving antenna is unable to couple electromagnetically with another particle by the plasmonic mechanism. If the SP damping rate $|\omega''_{l=1}(R)|$ is much larger than $\gamma^{diss}$, a plasmonic particle of the corresponding size $R$ is a good radiating antenna at the resonance frequency $\omega'_l(R)$ and is able to enhance the EM field in far and near field region.

Using intrinsic plasmon size characteristics $\omega'_l(R)$ and $\omega''_l(R)$ it is possible to control an operation resonance frequency of plasmonic nanoantenna and change the operation range from spectrally broad to spectrally narrow and vice versa (large or small $\omega''_l(R)$). It is also possible to switch between the particle receiving (enhanced absorption) and emitting (enhanced scattering) abilities in the qualitatively controlled manner. Changing the polarization geometry (Section 3.2) it is possible to effectively separate the dipole or quadrupole plasmon radiation from all the non-plasmonic contributions to the scattered light.

**Acknowledgment.** We would like to acknowledge support of this work by Polish Ministry of Education and Science under Grant No N N202 126837.